\documentclass{article}

\usepackage{amssymb}
\usepackage{amsfonts}
\usepackage{amsmath}
\usepackage{bm}

\newcommand{\la}{\langle}
\newcommand{\ra}{\rangle}
\newcommand{\nn}{\nonumber}

\newcommand{\beq}{\begin{equation}}
\newcommand{\eeq}{\end{equation}}
\newcommand{\beqy}{\begin{eqnarray}}
\newcommand{\eeqy}{\end{eqnarray}}
\newcommand{\beqyn}{\begin{eqnarray*}}
\newcommand{\eeqyn}{\end{eqnarray*}}

\newcommand{\bc}{\begin{center}}
\newcommand{\ec}{\end{center}}

\begin{document}

\bc THE TRANSVERSE ANGULAR MOMENTUM SUM RULE \ec

\bc Elliot.LEADER \ec

\bc High Energy Physics, Imperial College London,\\
London, SW7 2AZ, UK\\
E-mail: e.leader@imperial.ac.uk \ec

\begin{abstract}
We explain the origin of the controversy about the existence of a
transverse angular momentum sum rule, and show that it stems from
utilizing  an incorrect result in the literature, concerning the
expression for the expectation values of the angular momentum
operators. We demonstrate a new, short and direct way of obtaining
correct expressions for these expectation values, from which a
perfectly good transverse angular momentum sum rule can be
deduced. We also introduce a new classification of sum rules into
\emph{primary} and \emph{secondary} types. In the former all terms
occurring in the sum rule can be measured experimentally; in the
latter some terms cannot be measured experimentally.
\end{abstract}

\section{Background}
Shore and White's claim \cite{SW} that $a_0$ (which in the simple
parton model is equal to the contribution to the nucleon's angular
momentum arising from the quark spins) does \emph{not} contribute
to the nucleon's angular momentum, surprised us. Their analysis
was based on a classical paper of Jaffe and Manohar \cite{JM}, who
stressed the subtleties and warned that 'a careful limiting
procedure has to be introduced'. Trying to understand this we
became convinced that despite all the care, there are flaws. With
the J-M result one cannot have a sum rule for a transversely
polarized nucleon. With the correct version \cite{BLT} one can!

\section{Why the problem is non-trivial}
 What is the aim? We consider a nucleon with 4-momentum $p^{\,\mu}$ and covariant spin
vector $\textit{S}^{\,\mu}$ corresponding to some specification of
its spin state e.g. helicity, transversity or spin along the
Z-axis i.e. a nucleon in state $|p , \textit{S} \ra $. We require
an expression for the expectation value of the angular momentum in
this state i.e. for $ \la p ,\textit{S} |\bm{J} |p, \textit{S} \ra
$ i.e. we require an expression in terms of $p$ and $\textit{S}$.
This can then be used to relate the expectation value of $ \bm{J}$
for the nucleon to the angular momentum carried by its
constituents.
 \subsection{The traditional approach}
 In every field theory there is an expression for the angular momentum density operator.
The angular momentum operator $ \bm{J}$ is then an integral over
all space of this density. Typically the angular  momentum density
involves the energy-momentum tensor density $ T^{\mu\nu}(x)$ in
the form e.g.
 \beq \bm{J}_z = \bm{J}^3 = \int dV[xT^{02}(\bm{x})
-yT^{01}(\bm{x}) ] \nn \eeq Consider the piece $T^{02}(\bm{x})$.
It is a local operator, so by translational invariance of the
theory \beq T^{02}(\bm{x}) =e^{i\bm{P}.\bm{x}} T^{02}(0)
e^{-i\bm{P}.\bm{x}} \nn \eeq where $ \bm{P}$ are the linear
momentum operators i.e. the generators
 of translations.
  Now the nucleon is in an eigenstate of momentum, so $\bm{P}$
acting on it just becomes $\bm{p}$. The numbers $
e^{i\bm{p}.\bm{x}} e^{-i\bm{p}.\bm{x}}$ cancel out and we are left
with: \bc $ \int dV x \, \la p ,\textit{S}\, |\, T^{02}(0) \,| p,
\textit{S} \, \ra $ \ec The matrix element is independent of $x$
so we are faced with $ \int dV x $  $ =\infty$ ? or $ = 0$ ?
Totally ambiguous!
 The problem is an old one: In ordinary QM plane wave states
give infinities. The solution is an old one: Build a wave packet,
a superposition of physical plane wave states.
 Now Jaffe and Manohar are generally very careful, but
 nonetheless there are errors in their derivation. They end up
 with the following expression for the matrix elements of the
 angular momentum operator:
\beqyn
 \lefteqn{\la\la \bm{p}, \bm{s} | \bm{J}_i | \bm{p}, \bm{s}
\ra\ra_{JM} = } \\
&& \frac{1}{4mp_0}\left[ (3p^2_0 -m^2)\bm{s}_i -\frac{3p_0 + m
}{p_0 + m}(\bm{p}.\bm{s}) \bm{p}_i \right] \nn \eeqyn where $p^\mu
= (p^0, \bm{p})$ and $\bm{s}_i$ are the components of the rest
frame spin vector. Recall that the parton picture is supposed to
be valid when the nucleon is viewed in a frame where it is moving
very fast. In other words to derive a sum rule involving partons
we must take the limit $ p^0\rightarrow\infty $.
 If we consider \emph{longitudinal} spin i.e $\bm{p}\, //
\,\bm{s}$ one obtains: \beq \label{eq:JM} \la\la \bm{p}, \bm{s} |
\bm{J}_i | \bm{p}, \bm{s} \ra\ra_{JM} = \frac{1}{2}\bm{s}_i  \eeq

and there is no problem. But for transverse polarization one gets:
\beq \label{eq:JMTr}
 \la\la \bm{p}, \bm{s} | \bm{J}_i | \bm{p}, \bm{s}
\ra\ra_{JM} =
 \frac{1}{4mp_0}\left[ (3p^2_0 -m^2)\bm{s}_i \right]  \eeq

which $\rightarrow\infty $ as $p_0\rightarrow\infty$, so no
 sum rule is possible.
We will see in  a moment that the result for transverse spin is
 incorrect.
 The
J-M reaction to our criticism was very gracious and positive!

 ``Better late than never. Aneesh and I
finally found ourselves in the same place with the time to review
the issues you raised by email and in your recent paper. We agree
that there is an error in our eq. (6.9). It came from treating the
quantity u(p', s )u(p, s) with insufficient care. Thanks for
taking care and finding this mistake. It's good to get it cleared
up. I have to add that I found your paper rather difficult to
read. There is quite a bit of stuff that gets in the way of the
relatively simple error..........."

\subsection{A new approach}
It is simple. It is short. It works for any spin. Previous methods
only work for spin 1/2.
 We know how rotations affect states. If $ | \bm{p},m \ra$ is a
state with momentum $ \bm{p}$ and spin projection $m$ in the rest
frame of the particle, and if $ \hat{R}_z(\beta) $ is the operator
for a rotation $\beta$ about $OZ$, then
 \beq \hat{R}_z(\beta)| \bm{p},m \ra = |\bm{R}_z(\beta)\bm{p} ,
m' \ra \, D_{m'm}^s[R_z(\beta)] \label{eq:Rot}  \eeq where the
$D_{m'm}^s $  are the standard rotation matrices for spin $s$. But
rotations are generated by the angular momentum operators! i.e.
 \beq
\hat{R}_i(\beta) = e^{-i\beta\bm{J}_i} \nn \eeq so that
 \beq \bm{J}_i = i\frac{d}{d\beta}\hat{R}_i(\beta)\big|_{\beta=0}
\nn \eeq
 From Eq. (\ref{eq:Rot}) we know what the matrix element of $
\hat{R}_i(\beta) $ looks like. So we simply differentiate and put
$\beta = 0$. Thus we have \beq \la \bm{p}', m' |  \bm{J}_i |
\bm{p}, m \ra = i\frac{\partial}{\partial\beta}\,\la \bm{p}', m' |
R_i(\beta) | \bm{p}, m \ra|_{\beta=0}\nn \eeq One technical point:
you have to know that the derivative of the rotation matrix for
spin $s$ at $\beta=0$ is just the spin matrix for that spin. e.g.
for spin 1/2 just $\sigma_i/2 $.
\section{Comparison of results}
For the expectation values we find, for \emph{any} spin
configuration (longitudinal, transverse etc) the remarkably simple
result (suppressing a delta-function term): \beq \la\la \bm{p},
\bm{s} | \bm{J}_i | \bm{p}, \bm{s} \ra\ra = \frac{1}{2}\bm{s}_i
\nn \eeq This agrees precisely with the JM result for longitudinal
spin Eq. (\ref{eq:JM}).
 But for transverse polarization our result differs from the JM Eq.
 (\ref{eq:JMTr}),
which implied no possibility of a transverse sum rule. With our
correct result there is no fundamental distinction
 between the transverse and longitudinal cases.
 \section{Sum rules}
 Consider a nucleon moving along $OZ$ with momentum $\bm{p}$ and spin projection
$m$ along $OZ$. We expand the nucleon state as a superposition of
$n$-parton Fock states.
\begin{eqnarray}
 | \bm{p}, m \rangle = \sum_n
 \sum_{\{\sigma\}} \int d^3\bm{k}_1
 \dots d^3\bm{k}_n
\, \psi_{\bm{p}, m} (\bm{k}_1,\sigma_1, ... \bm{k}_n,
 \sigma_n)\nonumber \\
\times  \, \delta^{(3)}(\bm{p} - \bm{k}_1 ... - \bm{k}_n)
 \, |\bm{k}_1,\sigma_1,  ... \bm{k}_n, \sigma_n \rangle.
\nn \end{eqnarray}
 where $\sigma_j$ labels the spin state of the parton, either a projection along
 $OZ$ for quarks, or helicity for gluons.

 There
are two independent cases:

(a) \emph{Longitudinal polarization} i.e. the nucleon rest frame
spin vector $\bm{s}$ is along $OZ$. The sum rule for $\bm{J}_z$
yields the well known result \beq 1/2 = 1/2 \,\Delta \Sigma +
\Delta G +
 \langle L_{z}^q \rangle + \langle L_{z}^G \rangle
\label{eq:long} \eeq

(b) \emph{Transverse polarization} i.e. $\bm{s}= \bm{s}_T $ where
$\bm{s}_T \perp\bm{p} $. The sum rule for $\bm{J}_x$ or $\bm{J}_y$
yields a a \emph{new} sum rule

 \beq 1/2 = 1/2 \, \sum_{q, \,\bar q }\, \int dx \, \Delta _T q
(x) + \sum_{q, \, \bar q, \, G }\langle L_{\bm{s}_T} \rangle
\label{eq:trans} \eeq

 Here $L_{\bm{s}_T}$ is the component of
$\bm{L}$ along $\bm{s}_T$.

The structure functions $\Delta_T q^a (x) \equiv h^q_1(x)$ are
known as the quark transversity or transverse spin distributions
in the nucleon. As mentioned no such parton model sum rule is
possible with the J-M formula for the expectation value of
$\bm{J}_i$ because
 for $i=x,y$ the matrix elements diverge as $p \rightarrow
 \infty$.

 It is absolutely crucial to note that the sum rule Eq. (\ref{eq:trans}) involves a
\emph{sum} of quark and antiquark densities. Not realizing this
has led to some misunderstandings.

The \emph{tensor charge} of the nucleon involves the
\emph{difference} of the first moments of quark and antiquark
contributions. Thus the transverse spin sum rule, although it
involves the transverse spin or transversity quark and antiquark
densities, does \emph{not} involve the nucleon's tensor charge.
The tensor charge operator is \emph{not} related to the angular
momentum.

 The structure functions $\Delta_T q (x) \equiv
h^q_1(x)$ are most directly measured in doubly polarized Drell-Yan
reactions

\[ p(\bm{s}_T) + p(\bm{s}_T) \rightarrow l^{+} + l^{-} + X \]

where the asymmetry is proportional to

\beq
 \sum_f e^2_f [\Delta_T q_f (x_1) \Delta_T \bar{q}_f (x_2) + (1 \leftrightarrow 2)].
\nn \eeq

 They can also be determined from the asymmetry in semi-inclusive
hadron-hadron interactions like

\[ p + p\,(\bm{s}_T) \to H + X \]

 where $H$ is a detected hadron, typically a pion, and in
 semi-imclusive lepton-hadron reactions (SIDIS) with a transversely polarized target,  like

\[\ell + p\,(\bm{s}_T) \to \ell + H + X.\]

The problem here is that in these semi-inclusive reactions
$\Delta_T q (x)$ always occurs multiplied by the Collins
fragmentation function, about which we are only at present
gathering information.

\section{A new classification of sum rules}
Part of the reason that there are claims and counter-claims about
the existence of certain sum rules is that different people have a
different interpretation as to what a sum rule really implies. To
clarify this we propose a new classification into \emph{primary}
and \emph{secondary} sum rules.
\begin{itemize}

\item A \emph{primary} sum rule is one in which every term occurring
in the sum rule can be measured experimentally.  If the derivation
of the sum rule is rigorous and if it fails experimentally, one
can conclude that the theory behind it is incorrect. Examples are
the Bjorken sum rule \cite{Bj}, the ELT sum rule \cite{ELT} and
the Ji sum rule \cite{Ji}.
\item A \emph{secondary} sum rule is one in which not  every term
occurring can be measured experimentally. Examples are Eq.
(\ref{eq:long}) and Eq. (\ref{eq:trans}), where we do not know how
to measure the orbital angular momentum terms experimentally.
Consequently a secondary sum rule can't test the validity of a
theory, but this does not mean the sum rule is vacuous. It can
tell us about the terms which we cannot measure, and that can be
of value in model building or in understanding the structure of
say the nucleon. Do not forget that the renaissance of spin
dependent deep inelastic scattering, both theory and experiment,
is a direct consequence of using a secondary sum rule i.e. Eq.
(\ref{eq:long}) to proclaim the existence of a  ``spin crisis in
the parton model" \cite{AL}.
\end{itemize}

Of course the above is an idealization. I do not know of a single
case where literally everything is measurable. So in the Bjorken
and ELT sum rules one has to extrapolate $g_1(x)$ and $x[g_1(x)+
2g_2(x)]$ respectively to $x=0$, and in the Ji case one must
extrapolate $ E(x,\xi,\Delta ^2)$ to $\Delta^2 =0$. Nonetheless I
think the classification is useful.

\section{Conclusions}
In order to derive angular momentum sum rules we need an
expression for the matrix elements of the angular momentum
operators $\bm{J}$ in terms of the momentum $\bm{p}$ and spin
$\bm{s} $ of the particle. Such matrix elements are divergent and
ambiguous in the traditional approach. The infinities and
ambiguities can be handled using wave packets, but the
calculations are long and unwieldy. and the results, in some
classic papers, are incorrect for a transversely polarized
nucleon. Consequently it was claimed that no  angular momentum
rule was possible for a transversely polarized nucleon.

We have found a simple, direct method for evaluating these matrix
elements, which is free of infinities and ambiguities. It uses the
facts that we know how states transform under rotations, and that
the rotation operators are exponentials of the generators of
rotations i.e. of the angular momentum operators. It leads quickly
and relatively painlessly to correct results.

 The great success of the correct approach is that it allows the
derivation of a sum rule also for transversely polarized nucleons.

Finally, we have proposed a classification of sum rules into
\emph{primary} and \emph{secondary} sum rules, according to
whether all, or not all, the terms in a sum rule can be measured
experimentally. Whereas the former could, in principle, disprove a
theory, the latter can only give us information about quantities
which we cannot measure directly. \emph{Both} the longitudinal and
the transverse angular momentum sum rules are secondary.
\section*{Ackmowledgements}
This work was carried out in collaboration with Ben Bakker, Vrije
Universiteit, Amsterdam and Larry Trueman, Brookhaven National
Laboratory, USA.

\end{document}